\documentclass[intlimits,twoside,a4paper]{article}

\usepackage{amsmath,amssymb}
\usepackage{graphicx}
\usepackage{wrapfig}
\usepackage{siunitx}
\usepackage{cite}

\usepackage[T2A]{fontenc}
\usepackage[cp1251]{inputenc}

\usepackage[eqsecnum]{cmpj2}

\issue{2012}{15}{4}{43001}

\doinumber{10.5488/CMP.15.43001}
\title[Marginal dimensions for multicritical phase transitions]%
{Marginal dimensions for multicritical phase transitions%
}
\author[M. Dudka \textsl{et al.}]{M. Dudka\refaddr{label1},
        R. Folk\refaddr{label2}, Yu. Holovatch\refaddr{label1}, G. Moser\refaddr{label3}}
\addresses{
\addr{label1} Institute for Condensed Matter Physics of the National Academy of Sciences of Ukraine,\\
1 Svientsitskii St., 79011 Lviv, Ukraine
\addr{label2} Institut f\"ur Theoretische Physik, Johannes Kepler
Universit\"at Linz, A--4040, Linz, Austria
\addr{label3} Fachbereich f\"ur  Materialforschung und Physik, Univerit\"at
Salzburg, A--5020 Salzburg, Austria}
\date{Received June 18, 2012}

\authorcopyright{M. Dudka, R. Folk, Yu. Holovatch, G. Moser, 2012}

\begin{document}

\maketitle

\begin{abstract}
The field-theoretical model describing multicritical phenomena with
two coupled order parameters with $n_{||}$ and $n_\perp$ components
and of  $O(n_{\|}){\oplus} O({n_{\perp}}\!)$ symmetry is considered.
Conditions for realization of different types of multicritical
behaviour are studied within the field-theoretical renormalization
group approach. Surfaces separating stability regions for certain
types of multicritical behaviour in parametric space of order
parameter dimensions and space dimension $d$ are calculated using
the two-loop renormalization group functions. Series for the order
parameter marginal dimensions that control the crossover between
different universality classes are extracted up to the fourth order
in $\varepsilon=4-d$ and to the fifth order in a
pseudo-$\varepsilon$ parameter using the known high-order perturbative expansions for
isotropic and cubic models. Special attention is paid to a particular
case of $O(1){\oplus} O(2)$ symmetric model relevant for description of anisotropic antiferromagnets in an external magnetic field.
\pacs 05.50.+q, 64.60.ae
\end{abstract}

\section{Introduction}
The concept of universality plays a paradigmatic role in the modern
statistical physics. Accordingly, continuous  phase transitions can
be grouped into universality classes (see, e. g. \cite{domb}). Systems
within the same universality class are characterized by the same set
of critical exponents governing the scaling behaviour of their
thermodynamical functions. Therefore, one of the aims of a
theoretical description of a system  is to establish its
universality class.

\looseness=-1In the theory of critical phenomena it is standard now to use
methods  of field theoretical renormalization group (RG)
\cite{rgbooks,rgbooks_2,rgbooks_3,rgbooks_4}. Within these methods,  a stable fixed point (FP)
corresponds to the universality class. For systems with  complex internal
symmetries described by $\phi^4$  theories with several couplings,
several different nontrivial FPs may exist. Depending on global
parameters of a system, these FPs can interchange their stability
causing the system to trigger from one universality class to another.
The lack of a stable FP  can even mean that a continuous phase order
transition is transformed into a discontinuous. These global
parameters (that effect the FP stability) are spatial dimension $d$ and
the dimension $n$ of the order parameter (OP). In the $n$--$d$-space,
the regions of stable FPs are separated by borders and the $n(d)$
curves define the OP {\em marginal dimensions} that control
the crossover between different universality classes.

In this paper we are interested in  the stability borders and
marginal dimensions for a model with two coupled OP fields,
namely, the model with $O(n_{\|})\oplus O(n_{\perp})$ symmetry
\cite{NelsonKosterlitzFisher,AharonyBruce,KosterlitzNelsonFisher}.
Such a model  describes, amongst other systems \cite{Aharony_st},
anisotropic antiferromagnets in an external magnetic field
\cite{antiferromagnets,antiferromagnets_2,antiferromagnets_3,antiferromagnets_4,antiferromagnets_5,antiferromagnets_6,Folk08a}.

Conditions for realization of different types of multicritical
behaviour, that are defined by the relation between the dimensions of the OPs
$n_{\|}$, $n_{\perp}$, were obtained already in the first nontrivial
approximation of the field-theoretical RG for $d<4$
\cite{KosterlitzNelsonFisher,Aharony1976,Lyuksutov}.  They
determine the stability regions in the parametric $n_{\|}-n_{\perp}$
plane for three FPs: {\em isotropic Heisenberg} FP of
$O(n_{\|}{+}n_{\perp})$ symmetry, {\em decoupled} FP at which OPs
are ordering separately, and {\em biconical} FP. The two-loop
studies in $d=3$ show qualitatively similar results
\cite{Prudnikov98,Folk08a}, although  significantly changing the quantitative picture in
$n_{\|}-n_{\perp}$ plane.  Five-loop results for
three-dimensional $O(n_{\|})\oplus O(n_{\perp})$ model
\cite{Calabrese03}  confirm the obtained picture, producing only slight
corrections.

Since the previous studies of multicritical behaviour in the
$O(n_{\|})\oplus O(n_{\perp})$ system concentrated on $d=3$ case,
 in this paper we consider the dependence of marginal dimensions of
$O(n_{\|})\oplus O(n_{\perp})$ model on space dimension $d$. Our
motivation is caused by the fact that even a small change in $d$ can
produce crucial effects on the critical behaviour, in particular,
changing the universality class of a system. The rest of the paper is
organized as follows: In section~\ref{II} we present the
$O(n_{\|})\oplus O(n_{\perp})$ model and its RG description. Then,
our aim is to analyse the conditions for realizing different
scenarios of multicritical behaviour. In section~\ref{III}, we present the results obtained
within the two-loop approximation based on the $\varepsilon$-expansion
as well as on the fixed $d$ approach. We devote the next
section~\ref{IV} to the results for marginal dimensions of
$O(n_{\|})\oplus O(n_{\perp})$ in higher  order approximations. We
end the paper with section~\ref{V} where  our conclusions are presented.

\section{The model and RG picture of its multicritical phenomena}\label{II}

The  model with $O(n_{\|})\oplus O(n_{\perp})$ symmetry can be
obtained from the well-known $O(n)$-symmetrical model
\cite{Brezin1974}, splitting its $n$-component OP $\vec{\phi}_{0}$
into two: $\vec{\phi}_{\perp 0}$ and $\vec{\phi}_{\| 0}$  that
act in orthogonal  subspaces with dimensions $n_{\|}$ and
$n_{\perp}$, respectively ($n_{\|}+n_{\perp}=n$):
\begin{equation}\label{norder_parameter}
\vec{\phi}_0=\left(\begin{array}{c} \vec{\phi}_{\perp 0} \\  \vec{\phi}_{\| 0}
\end{array}\right) .
\end{equation}
Then, separating the Ginsburg-Landau-Wilson functional of $O(n)$ symmetry
one can present the effective Hamiltonian of the $O(n_{\|})\oplus
O(n_{\perp})$ model in the form:
\begin{eqnarray}\label{hbicrit}
{\cal H}_{\mathrm{Bi}}\!=\!\int\!
\rd^dx\Bigg[\frac{1}{2}\mathring{r}_\perp\vec{\phi}_{\perp 0}
\cdot\vec{\phi}_{\perp
0}+\frac{1}{2}\sum_{i=1}^{d}\nabla_i\vec{\phi}_{\perp
0}\cdot \nabla_i\vec{\phi}_{\perp 0}
+\frac{1}{2}\mathring{r}_\|\vec{\phi}_{\| 0} \cdot\vec{\phi}_{\|
0}+\frac{1}{2}\sum_{i=1}^{d}\nabla_i\vec{\phi}_{\| 0}\cdot
\nabla_i\vec{\phi}_{\| 0}  \nonumber \\
+\frac{\mathring{u}_\perp}{4!}\Big(\vec{\phi}_{\perp
0}\cdot\vec{\phi}_{\perp 0}\Big)^2
+\frac{\mathring{u}_\|}{4!}\Big(\vec{\phi}_{\|
0}\cdot\vec{\phi}_{\| 0}\Big)^2
+\frac{2\mathring{u}_\times}{4!}\Big(\vec{\phi}_{\perp
0}\cdot\vec{\phi}_{\perp 0}\Big) \Big(\vec{\phi}_{\|
0}\cdot\vec{\phi}_{\| 0}\Big) \Bigg] \ ,
\end{eqnarray}
where three couplings $\mathring{u}_\|$, $\mathring{u}_\perp$ and
$\mathring{u}_\times$ should be introduced instead  of the only one
in the $O(n)$ symmetric model, and $\mathring{r}_\perp$ and
$\mathring{r}_\|$  are connected with the temperature distance to
the critical line for $\vec{\phi}_{\perp 0}$ and $\vec{\phi}_{\| 0}$,
correspondingly.

The first mean-field analysis of the model with two coupled OPs was
performed in order to describe the supersolids \cite{LiuFisher} (see
also \cite{Lyuksutov}). It shows that the character of the
multicritical point in such a phase diagram depends on the sign of
$\mathring{u}_{\perp}\mathring{u}_{\|}-\mathring{u}^2_{\times}$. For
a positive sign, a tetracritical point is realized, while for a negative
sign, it is a  bicritical point. Going beyond the mean field theory,
fluctuations should be taken into account. This is achieved by the
field-theoretical RG approach \cite{rgbooks}, in which the
large-scale behaviour of the system is connected with the stable FP
of the  RG transformations. The transformation of the fourth order
couplings $\{\mathring{u}\}$ in (\ref{hbicrit}) under
renormalization is described by $\beta$-functions.

The $\beta$-functions for $O(n_{\|})\oplus O(n_{\perp})$ model were
known in a one-loop approximation~\cite{KosterlitzNelsonFisher}. The
next order approximation  has been found  in the massive
\cite{Prudnikov98} as well as in the minimal subtraction RG schemes
\cite{Folk08a}. In the minimal subtraction scheme, the
$\beta$-functions were  also calculated in the five-loop
approximation \cite{Calabrese03}, although explicit expressions were
presented only for $O(3)\oplus O(2)$ symmetry
\cite{Hasenbusch05}. Here, we work with $\beta$-functions obtained
in two-loop order \cite{Folk08a} within the minimal subtraction RG
scheme \cite{tHooft72,tHooft72_2}:
\begin{eqnarray}
\label{2} \beta_{u_\perp}&=& -\varepsilon  u_\perp +
\frac{(n_\perp+8)}{6} u_\perp^2 + \frac{n_\|}{6}u_\times^2 -
\frac{(3n_\perp + 14)}{12} u_\perp^3 - \frac{5n_\|}{36}u_\perp
u_\times^2
- \frac{n_\|}{9}u_\times^3\,,\\
\label{3} \beta_{u_\times} & = & -\varepsilon  u_\times +
\frac{(n_\perp+2)}{6}u_\perp u_\times + \frac{(n_\|+2)}{6}u_\times
u_\| + \frac{2}{3}u_\times^2 - \frac{(n_\perp+n_\|+16)}{72}
u_\times^3 \nonumber  \\
&&{}- \frac{(n_\perp+2)}{6}u_\times^2 u_\perp  - \frac{(n_\|+2)}{6}
u_\times^2  u_\| - \frac{5(n_\perp+2)}{72}
 u_\perp^2 u_\times  - \frac{5(n_ \|+2)}{72}u_\times u_\|^2\,, \\
\label{4} \beta_{u_\|} &=& -\varepsilon  u_\| + \frac{(n_\|+8)}{6}
u_\|^2 + \frac{n_\perp}{6}u_\times^2 - \frac{(3n_\| + 14)}{12}
u_\|^3 -\frac{5n_\perp}{36}u_\|u_\times^2 -
\frac{n_\perp}{9}u_\times^3\,.
\end{eqnarray}
Here, $\{u_\perp,u_\times,u_\|\}=\{u\}$ are renormalized couplings
and the space dimension $d$ enters the $\beta$-functions via
$\varepsilon=4-d$.

The FPs  $\{u^*\}$ of the RG transformation are found from the zeros
of the $\beta$-functions
\begin{equation} \label{6}
\beta_{u_i}(\{u^*\})=0
\end{equation}
with $i=\perp,\, \|, \, \times$. A stable FP possesses positive
eigenvalues $\omega_1$, $\omega_2$, $\omega_3$ (or their real parts)
of stability matrix $\partial \beta_i/\partial u_j$.

The stable FPs for $O(n_{\|})\oplus O(n_{\perp})$ are  already known
from the one-loop studies \cite{KosterlitzNelsonFisher}. For $d<4$
and for sufficiently low OP dimensions  satisfying
\begin{equation}\label{cond1}
n_{\perp}+n_{\|}<4,
\end{equation}
only the isotropic Heisenberg FP ${\cal H}$ of $O(n_{\perp}+n_{\|})$
symmetry with $\{u_\perp^*=u_\times^*=u_\|^*\}$ is stable.
When $n_{\perp}$ (or $n_{\|}$) increases breaking  (\ref{cond1}),
still with
\begin{equation}\label{cond2}
n_{\perp}n_{\|}+2(n_{\perp}+n_{\|})<32,
\end{equation}
 FP ${\cal H}$ interchanges its stability with biconical FP ${\cal B}$ $\{u_\perp^*\neq
u_\times^*\neq u_\|^*\}$.  For values of $n_{\perp}$ and $n_{\|}$
that are above the condition (\ref{cond2}),   FP ${\cal B}$ looses its
stability, while the decoupled FP ${\cal D}$ $\{u_\perp^*\neq 0,
u_\times^*=0, u_\|^*\neq 0\}$ becomes stable. According to these
one-loop results, the multicritical behaviour of the $O(1)\oplus O(2)$ model
is governed by FP ${\cal H}$ (connected with bicriticality) for all
space dimensions $d<4$. However, within the higher order
calculations, the stability of FPs depends not only on $n_\|,\,
n_\perp$ but also on $d$. Using resummation procedures for the
two-loop RG functions at $d=3$, one can show that the conditions of
the FPs stability (\ref{cond1}) and (\ref{cond2}) are drastically
shifted to smaller values of OP components
\cite{Prudnikov98,Folk08a}. In particular, in the case $n_\|=1$,
$n_\perp=2$ FP ${\mathcal B}$ (connected with tetracriticality)
appears to be stable in a two loop order \cite{Folk08a}. Resummation
of   higher orders $\varepsilon$-expansion \cite{Calabrese03} does
not change this result.

\section{Stability border-surfaces  within a two-loop order approximation} \label{III}

As  noted above, the stability of FPs ${\mathcal{D}}$,
${\mathcal{B}}$, ${\mathcal{H}}$ is dependent on three parameters
$n_\|$, $n_\perp$ and $d$. Therefore, the borders between regions
for which one or another FP is stable, form surfaces in the
parametric space $n_\|-n_\perp-d$: $f(n_\|,n_\perp,d)=0$. We call
them {\it border-surfaces} (BSs).

Two alternative ways are used in practice to analyze RG functions
and to get universal quantities, in particular, marginal dimensions.
In one approach, i.e., the $\varepsilon$-expansion, the solutions are
obtained as a series in $\varepsilon$ and then they are evaluated at the value
of interest (for instance, at $\varepsilon=1$ for $d=3$ theories).
Alternatively, one may fix the space dimension $d$  to a certain value
and  directly solve a system of non-linear equations obtaining the
FP coordinates numerically  \cite{Schloms}.  In the next two
subsections we use these approaches to obtain marginal dimensions of the
$O(n_\|)\oplus O(n_\perp)$ model within a two-loop RG
approximation.

\subsection{BSs from $\varepsilon$-expansion}

\begin{figure}[ht]
\begin{picture}(400,150)
\put(-5,10)
{\includegraphics[width=0.48\textwidth]{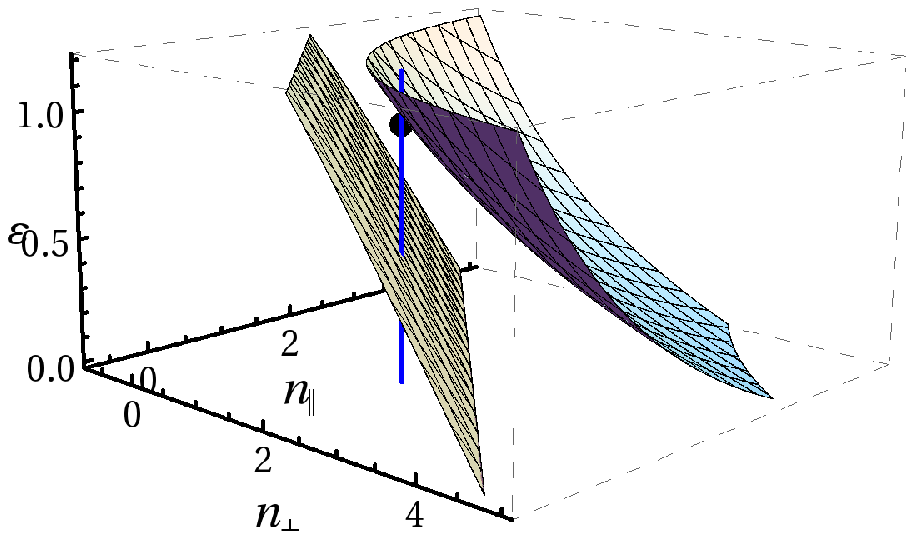}}
\put(220,10)
{\includegraphics[width=0.48\textwidth]{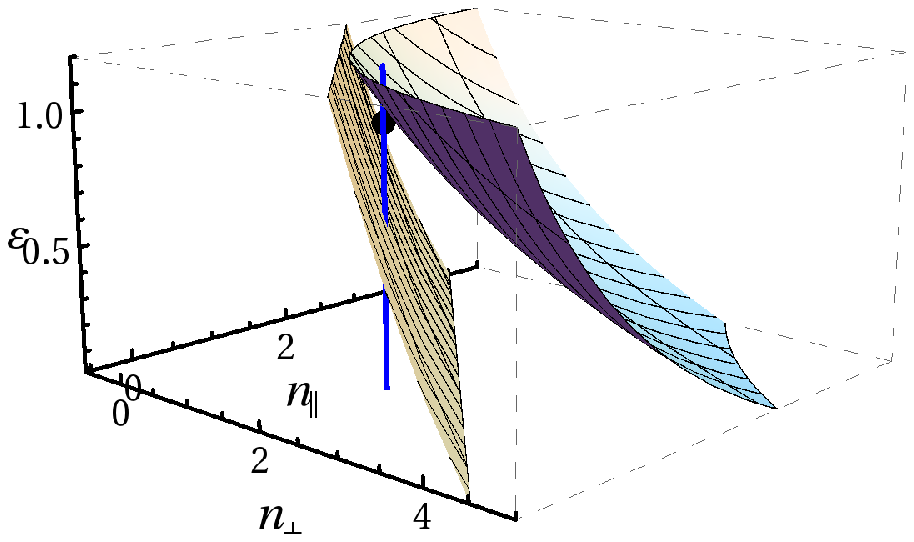} }
\put(100,0)
{(a)}
\put(325,0)
{(b)}
\end{picture}
\caption{(Color online)\label{epsilon} BSs between different universality classes
of the $O(n_{||})\oplus O(n_{\perp})$ model obtained (a) by applying an
$\varepsilon$-expansion  and (b) by using a resummation procedure  at a fixed
$d$  to two-loop RG functions. The left hand (lower) surface separates
the stability region of FP $\mathcal H$ (on the left from the
surface) and  FP $\mathcal B$ (on the right from the surface). The
right hand (upper) surface separates the stability regions of  FP
$\mathcal B$ (on the left from surface) and $\mathcal D$ (on the
right from surface). The vertical line shows the position of a
system  with $n_{\|}=1,\,n_{\perp}=2$. The disc on the line indicates
the position at $d=3$.}
\end{figure}

We start our analysis with establishing the border between the
stability regions of the decoupled FP $\mathcal{D}$ and the
biconical FP $\mathcal{B}$. As it was noted in \cite{Folk08a}, two of
the FP $\mathcal{D}$ stability exponents  correspond to the
stability exponent of the $O(n)$ model $\omega^{\mathcal{H}(n)}$:
$\omega^{\mathcal{D}}_1=\omega^{\mathcal{H}(n_{\|})}$,
$\omega^{\mathcal{D}}_3=\omega^{\mathcal{H}(n_{\perp})}$, while the
remaining one is defined by
\begin{equation} \label{omegas}
\omega^{\mathcal{D}}_2=\left.{\partial
\beta_{u_{\times}}}/{u_{\times}}\right|_{\mathcal{D}}\,.
\end{equation}
Since $\omega^{\mathcal{H}(n)}$  is always positive, only
$\omega^{\mathcal{D}}_2$ governs the stability of  the FP
$\mathcal{D}$, changing its sign depending on $n_{\|}$, $n_{\perp}$,
$d$. Therefore, the surface between stability regions of FPs
$\mathcal{D}$ and $\mathcal{B}$ can be extracted from the condition
of   (\ref{omegas}) vanishing. Substituting the
$\varepsilon$-expansion for the FP $\mathcal{D}$ coordinates into
(\ref{omegas}) one collects  terms up to $\varepsilon^2$ and sets
the result equal to zero:
\begin{equation}\label{eq1}
\varepsilon    \left[\frac{ (13 n_{\|}+44)
(n_{\|}+2)}{2(n_{\|}+8)^3}+\frac{ (n_{\perp}+2)
   (13 n_{\perp}+44)}{2(n_{\perp}+8)^3}\right]+
 \left[
 \frac{n_{\|}-4}{2(n_{\|}+8)}+\frac{n_{\perp}-4}{2(n_{\perp}+8)}\right]=0.
\end{equation}
This is analytically solved for $\varepsilon=\varepsilon(n_{||}
,n_\perp)$. The result is shown as the right hand surface in
figure~{\ref{epsilon}}~(a).

The BS  between the regions of stability of the FPs $\mathcal{B}$ and
$\mathcal{H}$ can be derived from the condition that FP
$\mathcal{H}$ changes its stability.  Only one of the three
eigenvalues of the stability matrix changes its sign in the region
considered. Calculating this eigenvalue  up to the $\varepsilon^2$
order we get the equation for the surface:
\begin{equation}
\frac{\left[-5 ({n_{\perp}+n_{\|}}+8)^2+66
({n_{\perp}+n_{\|}}+8)-360\right] \varepsilon
}{({n_{\perp}+n_{\|}}+8)^3}+\left(\frac{12}{{n_{\perp}+n_{\|}}+8}-1\right)=0.
\end{equation}
The surface is also shown in figure~{\ref{epsilon}}~(a) (the lower left hand
surface).

The limiting borderlines in the plane $\varepsilon=0$ ($d=4$), are
equivalent to the case when the one loop order inequalities
(\ref{cond1}), (\ref{cond2}) are transformed into equalities, from
which one obtains
\begin{equation}\label{1lalines}
n_\|^{\mathcal D}(n_\perp) = \frac{2(16-n_\perp)}{n_\perp+2}, \qquad
n_\|^{\mathcal H}(n_\perp) = - n_\perp + 4.
\end{equation}

The vertical line in figure~\ref{epsilon}~(a) presents a system with
$n_{\|}=2$, $n_{\perp}=1$, indicating which FP governs the multicritical
behavior of this system with the change of $\varepsilon$. Note that the
FP $\mathcal B$ is stable in the region $0.51\lesssim
\varepsilon\lesssim1.04$. We are  interested in this case, since it
describes anisotropic ferro- and antiferromagnets in space dimension $d=3$.

\subsection{BSs from resummed $\beta$-functions}\label{III2}

Another way  to obtain the BSs, is to calculate them from the
$\beta$-functions  (\ref{2})--(\ref{4}) fixing $d$ at certain values.
Since the RG expansions have divergent \cite{rgbooks} nature, the
special resummation techniques are needed  to get convergent results
\cite{Holovatch02}. The two-loop $\beta$-functions
(\ref{2})--(\ref{4}) $\beta=\beta_{u_i}$ have a form of  polynomials
in renormalized couplings:
\begin{equation}\label{ap1}
\beta(\{u\})=\sum_{1\leqslant i,j,k\leqslant 3}c_{ijk}u_\perp^i u_\|^j
u_\times^k\,.
\end{equation}
We first represent (\ref{ap1}) in the form of a resolvent series
\cite{Watson74} in one auxiliary variable $t$:
\begin{equation}\label{ap2}
F(\{u\},t)=\sum_{1\leqslant i,j,k\leqslant 3}c_{ijk}u_\perp^i u_\|^j u_\times^k t^{i+j+k-1} =
\sum_{0\leqslant \alpha\leqslant 2} a_\alpha(\{u\},\{c\})t^\alpha,
\end{equation}
where the expansion coefficients $a_\alpha$ in (\ref{ap2}) explicitly
depend on the couplings and on the coefficients $c_{ijk}$ (\ref{ap1}).
Obviously, $F(\{u\},1)=\beta(\{u\})$. We resume the function
(\ref{ap2}) as a single variable function using the Pad\'e-Borel
technique \cite{Baker78} and writing its Borel image as:
\begin{equation}\label{ap3}
F^{\mathrm{B}}(t)= \sum_{0\leqslant \alpha\leqslant 2}{\frac{a_\alpha t^\alpha}{\alpha!}}\,.
\end{equation}
Analytical continuation of the function (\ref{ap3}) is achieved by
representing it in the form of a Pad\'e approximant \cite{Baker81}. In
our case, we use the diagonal  Pad\'e approximant [1/1]:
\begin{equation}\label{ap4}
F^{\mathrm{B}}(t) \simeq [1/1](t).
\end{equation}
Finally, the resummed function is obtained via an inverse Borel
transform:
\begin{equation}\label{ap5}
F^{\rm res}= \int_0^{\infty} [1/1](t) \re^{-t}.
\end{equation}

Applying the above procedure to the two-loop $\beta$-functions
(\ref{2})--(\ref{4}) at a fixed $d$ and  searching  for their FP
solutions with $u^*_\times=0$ and
together with expression (\ref{omegas}), where  $\omega_2^{\mathcal{D}}=0$, we find a BS, separating
the stability region of the FP ${\mathcal{D}}$ from the FP
${\mathcal{B}}$ stability region. It is depicted as the upper
surface in figure~\ref{epsilon}~(b). Searching for the FP solutions with
$u_\perp^*=u_\times^*=u_\|^*$ at which the  determinant of the
stability matrix  vanishes  we derive the BS between the stability
regions of the FPs ${\mathcal{H}}$ and ${\mathcal{B}}$. This is the
upper left hand surface in figure~{\ref{epsilon}}~(b).

It is technically difficult to extract the data from the resummed
function in the limit $\varepsilon\to 0$. Therefore, we present BSs
for $0.002\leqslant \varepsilon\leqslant 1.2$, and $n_{\|},\,n_{\perp}$ in the
range from --0.56 to 5. Limiting borderlines in the plane
$\varepsilon\to 0$ described by (\ref{1lalines}) give us the
one-loop (thin) borderlines  of figure~1 of \cite{Folk08a}, while
the intersections of the  surfaces with the plane $\varepsilon=1$
give the two-loop (thick) borderlines of  figure~1 of
\cite{Folk08a}.

\begin{figure}[ht]
\centerline{
\includegraphics[width=0.4\textwidth]{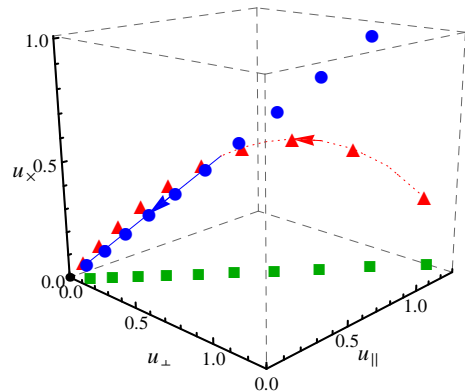}
}
\caption{(Color online) \label{trackfp} Locations of the FPs $\mathcal B$
(triangles), $\mathcal H$ (discs) and $\mathcal D$ (squares) for
$n_{\perp}=2,\, n_{\|}=1$ and $\varepsilon$ changing from 1 (end
right hand marks) to 0 with a step size of 0.1. Arrows show the direction
in which $\varepsilon$ decreases. Results are obtained using
resummation (\ref{ap2})--(\ref{ap5}). The line shows the track of a
stable FP. Dotted part of the line indicates that FP $\mathcal B$ is
stable, while the solid part indicates that FP $\mathcal H$ is
stable. Black disc at the origin indicates the Gaussian FP.}
\end{figure}
Similarly to what we did it in figure~\ref{epsilon}~(a), we present in
figure~\ref{epsilon}~(b) the line indicating the stability regions for FPs of
the $O(1)\oplus O(2)$ model. In this approximation for the
$\beta$-functions,  FP ${\mathcal B}$ is stable in the region
$0.66\lesssim\varepsilon\lesssim 1.06$, in particular at
$\varepsilon=1$ ($d=3$).  Let us check how the FP picture changes
along the line  with an increasing $d$. Varying the space dimension $d$
from 3 to 4 with stepsize $0.1$ we can observe the drift of FPs
$\mathcal B$, $\mathcal H$, $\mathcal D$ towards to the Gaussian FP.
The tracks are shown in figure~\ref{trackfp}  by triangles,
 discs and  squares indicating the change of locations of FPs
${\mathcal B}$,  ${\mathcal H}$ and   ${\mathcal D}$ in
$u_{\perp}-u_{||}-u_\times$ space   with the change of $d$.
Numerical values of the coordinates of these FPs  are listed in
table~\ref{tab}. FP ${\mathcal B}$ is stable up to the intersection
of traces of FP ${\mathcal B}$ and FP ${\mathcal H}$, which happens
at  $d\approx 3.34$, where it interchanges its stability with FP
${\mathcal H}$. Thus, only the FP ${\mathcal H}$ is stable starting
from the intersection point and up to the Gaussian FP. This would
mean that in higher space dimensions and at $n_{||}=1$,
$n_{\perp}=2$, the phase diagram contains a bicritical point
instead of a tetracritical point.
\begin{table*}
\caption{Coordinates of the FPs $\mathcal B$, $\mathcal H$, and
$\mathcal D$ at $n_{\|}=1$, $n_{\perp}=2$ depending on
$\varepsilon$ as found from the resumed two loop $\beta$-functions.
 \label{tab}}
\centering \tabcolsep=2mm
\begin{tabular}{|l|l|l|l|l|l|l|}
 \hline
   $\varepsilon$ & $\vphantom{1^{1^1}}$ $u_{\perp}^{{\mathcal B},*}$ & $u_{\|}^{{\mathcal B},*}$ & $u_{\times}^{{\mathcal B},*}$ & $u^{{\mathcal H},*}$
  & $u_{\perp}^{{\mathcal D},*}$ & $u_{\|}^{{\mathcal D},*}$ \\ \hline\hline
 1. & 1.1277 &  1.2874 & 0.3013 & 1.0016 & 1.1415 & 1.3146\\
 0.9 & 0.9112 & 1.0039 & 0.5273 & 0.8434 & 0.9569 & 1.0971\\
 0.8 & 0.7313 &  0.7739 & 0.5799 & 0.7026 & 0.7939 & 0.9063\\
 0.7 & 0.5834 &  0.5934 & 0.5518 & 0.5771 & 0.6496 &
 0.7386\\
 0.6 & 0.4596 &  0.4502 & 0.4863 & 0.4650 & 0.5214 &
 0.5906\\
 0.5 & 0.3543 &  0.3348 & 0.4048 & 0.3647 & 0.4075 &
 0.4600\\
 0.4 & 0.2634 &  0.2405 & 0.3183 & 0.2749 & 0.3061 &
 0.3444\\
 0.3 & 0.1843 &  0.1627 & 0.2323 & 0.1944 & 0.2158 &
 0.2420\\
 0.2 & 0.1149 &  0.0982 & 0.1497 & 0.1223 & 0.1354 & 0.1513
 \\
 0.1 & 0.0539 &  0.0446 & 0.0720 & 0.0578 & 0.0637 &
 0.0710\\
\hline
\end{tabular}
\end{table*}

\section{High loop order results  for marginal
dimensions}\label{IV}

The marginal dimensions of the $O(n_{\|})\oplus O(n_{\perp})$ models can
be defined based on the high order RG results for simpler
isotropic and cubic models. In particular, exact scaling arguments
\cite{Aharony1976} connect the FP ${\mathcal D}$ stability  with the
critical exponents of the $O(n_{\perp})$ and $O(n_{\|})$ models:
\begin{equation}\label{scalingargument}
\omega_2^{\mathcal
D}=-\frac{1}{2}\left[\frac{\alpha(n_{\perp})}{\nu(n_{\perp})}+\frac{\alpha(n_{\|})}{\nu(n_{\|})}\right] =d-\frac{1}{\nu(n_{\perp})}-\frac{1}{\nu(n_{\|})}\,,
\end{equation}
where $\alpha(n)$ and $\nu(n)$ are the heat capacity and correlation
length critical exponents of the $O(n)$ model.

As it was indicated in \cite{Calabrese03}, the stability of the
FP $\mathcal H$ is defined by the marginal dimension $n_{\mathrm{c}}$ of the
cubic model. Since in the FP $\mathcal H$, the RG functions depend
only on the combination $n=n_{\perp}+n_{\|}$, the resulting marginal
dimension can be presented in the form $n^{\mathcal
H}_{\perp}(n_{\|},\varepsilon)=n_{\mathrm{c}}(\varepsilon)-n_{\|}$.

In the following two subsections we present an analysis of the
marginal dimensions $n^{\mathcal D}_{\perp}(n_{\|},\varepsilon)$, \linebreak
$n^{\mathcal H}_{\perp}(n_{\|},\varepsilon)$ based on the
five-loop minimal subtraction series for the RG functions of
isotropic \cite{KleinertFrohlindeChetyrkinLarin1991} and cubic
models \cite{KleinertFrohlinde1995}, as well as for the case $d=3$ based on
the six-loop series for these models
\cite{Antonenko1995,Folk00} obtained within the massive scheme
\cite{Parisi,Parisi_2}.

\subsection{Five-loop $\varepsilon$-expansions for
marginal dimensions}

Let us start with the calculation of $n^{\mathcal
D}_{\perp}(n_{\|},\varepsilon)$. Substituting the five-loop
$\varepsilon$-expansions of the $O(n)$ theory
\cite{KleinertFrohlindeChetyrkinLarin1991} into
(\ref{scalingargument}) and  putting $\omega_2^{\mathcal D}$ equal
to zero, we get the equation for the BS. Keeping $n_{\|}$ as a
parameter and expanding in $\varepsilon$, we get $n^{\mathcal
D}_{\perp}(n_{\|},\varepsilon)$    in the following form:
\begin{eqnarray}
n_{\perp}(n_{\|},\varepsilon){\cdot}(n_{\|}+2)&=&{2
(16{-}n_{\|})}{-}{48}\varepsilon {+}8{\left[3 \zeta (3)
\left(n_{\|}^2{+}34 n_{\|}{+}100\right){+}\left(n_{\|}^2{+}58
n_{\|}{+}148\right)\right]}{ R_{n_\|}^2}\varepsilon^2\nonumber\\&&
{}{+}\Bigg\{{\left(11 n_{\|}^4-920 n_{\|}^3-528 n_{\|}^2+21376
n_{\|}+51584\right)}\big/3\Bigg. \nonumber\\
 &&{}- {4 \left(106592 + 64480 n_{\|}
+ 13548 n_{\|}^2 + 1258 n_{\|}^3 + 17 n_{\|}^4\right) \zeta (3)}\big/3
\nonumber\\&&{}+\Bigg.{\left[18 \left(100 + 34 n_{\|}+ n_{\|}^2\right) \zeta
(4)-40 \left(550 + 163 n_{\|} + 7 n_{\|}^2\right) \zeta
   (5)\big/3\right]}{    R_{n_\|}^{-2}}\Bigg\}{R_{n_\|}^4}\varepsilon^3\nonumber\\
   &&{}+\Bigg[
   \left(3 n_{\|}^6{+}170 n_{\|}^5{-}43120 n_{\|}^4{-}442864 n_{\|}^3{-}2069072 n_{\|}^2- 4512896 n_{\|}{-}3457280\right)\big/6\Bigg.\nonumber\\&&{}
  -   50  \left(550 + 163 n_{\|}+ 7 n_{\|}^2\right) \zeta (6)
   {   R_{n_\|}^{-4}}\big/3\nonumber
      \end{eqnarray}
   \begin{eqnarray}
   &&{}+{   \zeta (3) \left(1816192{+}
 5011904 n_{\|}{+} 3131936 n_{\|}^2{+}623376 n_{\|}^3{+}41740 n_{\|}^4{+}1486 n_{\|}^5{-}n_{\|}^6\right)
   }\big/3\nonumber\\&&{}-{
    4  \left(151424 +131552 n_{\|} +15728n_{\|}^2 -1250 n_{\|}^3 - 17 n_{\|}^4 - 5 n_{\|}^5\right)\zeta^2 (3)
   }{
   R_{n_\|}^{-1}}\big/3\nonumber\\
   &&{}-{\left(106592 + 64480 n_{\|} + 13548 n_{\|}^2 + 1258 n_{\|}^3 + 17n_{\|}^4\right) \zeta(4)
   }{
   R_{n_\|}^{-2}}\nonumber\\
   &&{}+{  \left(4822640 + 2331088 n_{\|}+ 416334n_{\|}^2 +  51745 n_{\|}^3+ 1103 n_{\|}^4\right) \zeta (5)}{
   R_{n_\|}^{-2}}\big/9\nonumber\\
   &&{}+\Bigg.
   {    49  \left(66320 + 31792n_{\|} + 5826 n_{\|}^2 + 535 n_{\|}^3 + 17 n_{\|}^4\right) \zeta (7)}{
   R_{n_\|}^{-2}}/2\Bigg]{R_{n_\|}^6}\varepsilon^4,
   \label{nd}
   \end{eqnarray}
where $R_n=(n+8)^{-1}$.
Expressions for certain physical values of $n_{\|}$ are less cumbersome:
\begin{eqnarray}
\label{n1} n_{\perp}(1,\varepsilon)&=&10.- 16.\varepsilon + 22.84224\varepsilon^2 -
 44.06758 \varepsilon^3 + 113.6428 \varepsilon^4,\\
 n_{\perp}(2,\varepsilon)&=&7. - 12.\varepsilon + 17.76523 \varepsilon^2 -
 34.18402 \varepsilon^3 + 84.07657 \varepsilon^4,\\
\label{n3} n_{\perp}(3,\varepsilon)&=&5.2 - 9.6 \varepsilon + 14.43837 \varepsilon^2 -
 28.00490 \varepsilon^3 + 67.23923 \varepsilon^4.
\end{eqnarray}
The obtained $\varepsilon$-expansion diverges, as it can be seen from the
growth of the expansion coefficients in (\ref{n1})--(\ref{n3}) as
well as it follows from the Pad\'e table \cite{Baker81} for
$n_{\perp}(1,1)$:
\begin{equation}\label{Pade-min-sub}
n_{\perp}(1,1)=\left(
\begin{array}{ccccc}
 10.0000 & 3.8462 & 3.4773 & 2.4576 & 9.6637 \\
 -6.0000 & 3.4092 & 3.9728 & 3.1128 & o \\
 16.8422 & 1.7981 & 2.8846 & o & o \\
 -27.2253 & 4.5288 &o & o & o \\
 86.4175 & o & o& o & o \\
\end{array}
\right).
\end{equation}
The element $MN$ of the table (\ref{Pade-min-sub}) is the value of
$n_{\perp}(1,1)$ given by the  $[M/N]$ Pad\'e approximant at
$\varepsilon=1$. Here and below, symbol $o$ denotes the approximants which can
not be constructed within the order of perturbation theory considered here.
Usually, the best convergence of the results is observed
along the main diagonal and the closest sub-diagonals of the Pad\'e table~\cite{Baker81}. However, it appears that  the value of
$n_{\perp}(1,1)$ given by the Pad\'e-aproximant [2/2] differs from
those given by [1/2] and [2/1] by an order of one, leading to an
uncertainty of the numerical estimate.

\begin{figure}[htbp]
\centerline{
\includegraphics[width=0.37\textwidth]{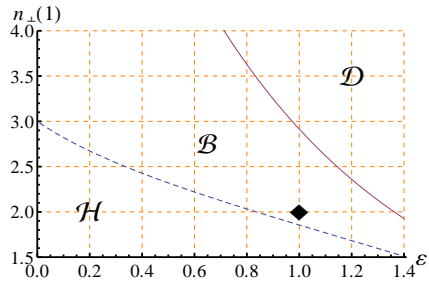}
}
\caption{(Color online) \label{5LA}The dependence of the marginal
dimensions $n^{\mathcal D}_{\perp}(1)$ (solid line) and $n^{\mathcal
H}_{\perp}(1)$ (dashed line) on $\varepsilon$. The results are obtained
based on the five-loop expansion for  isotropic and cubic models
using Pad\'e-Borel resummation  with [3/1] Pad\'e approximant (see
the text). The diamond denotes the location of the three-dimensional
$O(1)\oplus O(2)$ system.}
\end{figure}

To obtain a reliable estimate of $n_{\perp}(1,\varepsilon)$ we rely on the Pad\'e-Borel resummation described in subsection
\ref{III2}. We obtain a resolvent series by a substitution
$\varepsilon \to \varepsilon t$. For the obtained expression we
build the Borel-image, then approximate it by the [3/1] Pad\'e
approximant. Performing an integration of the inverse Borel transform,
we arrive at the result shown in figure~\ref{5LA} with a solid line.
In a similar way, we get $n^{\mathcal
H}_{\perp}(n_{\|},\varepsilon)$ using the available five-loop
$\varepsilon$-expansion for the marginal dimension of the cubic
model \cite{KleinertFrohlinde1995}:
\begin{eqnarray}
n_{\mathrm{c}}&=&4-2 \varepsilon+\left(-\frac{5}{12}+\frac{5 \zeta
(3)}{2}\right) \varepsilon ^2
   +\left(\frac{15 \zeta (4)}{8}-\frac{25 \zeta (5)}{3}+\frac{5 \zeta (3)}{8}-\frac{1}{72}\right) \varepsilon
   ^3\nonumber\\ &&{}+\left(\frac{15 \zeta (4)}{32}-\frac{125 \zeta
(6)}{12}+\frac{11515 \zeta (7)}{384}-\frac{3155 \zeta
(5)}{1728}-\frac{229 \zeta (3)^2}{144}+\frac{93 \zeta
   (3)}{128}-\frac{1}{384}\right) \varepsilon ^4.
\end{eqnarray}
The dependence of $n^{\mathcal
H}_{\perp}(1,\varepsilon)=n_{\mathrm{c}}(\varepsilon)-1$ on $\varepsilon$ is
obtained as above by the Pad\'e-Borel resummation with [3/1] Pad\'e
approximant. The result is shown in figure~\ref{5LA} with a dashed line.

As it can be seen from figure~\ref{5LA}, that FP ${\mathcal B}$ for
the $O(1)\oplus O(2)$ model is stable in the region
$0.84\lesssim\varepsilon\lesssim 1.36$. The value of $n_{\perp}=2$,
$d=3$ (denoted by a diamond) is located very close to the boundary
$n^{\mathcal H}_{\perp}(1,\varepsilon)$. Thus, one concludes a very
slow approach to the FP. Measurements in $O(1)\oplus O(2)$ systems
may show an effective critical behaviour with the exponents close to the
$O(3)$ case \cite{Folk08a}. Anyway, in recent Monte-Carlo
simulations of the Heisenberg ferromagnet with uniaxial exchange
anisotropy, only a bicritical point with Heisenberg symmetry was
obtained \cite{Selke2011}.

\subsection{Marginal dimension for of $d=3$ in a six-loop order}

Fixing the spatial dimension to  $d=3$, we can analyze $n^{\mathcal
D}_{\perp}(n_{\|})$ using pseudo-$\varepsilon$ expansions (for
details see \cite{LeGuillou80}). Introducing the
pseudo-$\varepsilon$ parameter $\tau$ into 6-loop RG functions of the massive scheme at a fixed $d=3$ \cite{Antonenko1995} for the $O(n)$
model, one can derive critical exponents in the form of
pseudo-$\varepsilon$ expansions and  substitute them into equation~(\ref{scalingargument}).  Similar to the former subsection, we
extract the pseudo-$\varepsilon$ expansion for $n^{\mathcal
D}_{\perp}(n_{\|})$ and present as an example
   \begin{eqnarray}\label{tau1}
   n_{\perp}(1)&=&10-10.66667 \tau+5.13069\tau ^2-2.30752\tau ^3+1.69527\tau ^4-1.98282\tau ^5,\\
\label{tau2}   n_{\perp}(2)&=&7-8\tau+3.99473\tau ^2-1.76429\tau ^3+1.140396\tau ^4-1.20818\tau ^5, \\
\label{tau3}   n_{\perp}(3)&=&5.2-6.4\tau+3.24817\tau ^2-1.46248\tau
^3+0.84832\tau ^4-0.84314\tau ^5.
      \end{eqnarray}
The pseudo-$\varepsilon$ expansions have better convergent
properties, as it is known from other studies
\cite{Dudka01,Folk00,Holovatch04}. This is  also seen from the
coefficients in the series (\ref{tau1})--(\ref{tau3}), as well as
from a comparison of the Pad\'e table (\ref{Pade-min-sub}) with the
one that follows from the pseudo-$\varepsilon$ expansion
(\ref{tau1}):
\begin{equation}
n_{\perp}(1)=\left(
\begin{array}{cccccc}
 10. & 4.8387 & 3.7156 & 3.2882 & 3.1541 & 3.0391 \\
 -0.6667 & 2.7977 & 2.8683 & 3.0805 & 2.0645 & o \\
 4.4640 & 2.8724 & 2.7743 & 2.9854 & o & o \\
 2.1565 & 3.1338 & 3.0073 & o & o & o \\
 3.8518 & 2.9379 & o & o & o & o \\
 1.8690 & o & o & o & o & o
\end{array}
\right).
\end{equation}

However, the convergence of the results might be spoiled if a pole in
the denominator of a Pad\'e approximant appears. We demonstrate this
below by the Pad\'e table for $n_{\perp}(3)$:
\begin{equation}
n_{\perp}(3)=\left(
\begin{array}{cccccc}
 5.2 & 2.3310& 1.6662& 1.3945 & 1.2670 & 1.1823 \\
 -1.2 & 0.9546 & 1.0319 & 1.1042 & 0.9528 &o \\
 2.0482 & 1.0397& ^{-0.4474} & 1.0623 & o & o \\
 0.5857 & 1.1226 & 1.0705 & o & o & o \\
 1.4340 & 1.0112 & o & o & o & o \\
 0.5909 & o & o & o & o & o
\end{array}
\right),
\end{equation}
where by small digits we indicate a result for the approximant [2/2] with a pole for
$\tau=0.944$.

From the Pad\'e-Borel procedure with $[4/1]$ approximant, we get:
$n^{\mathcal D}_{\perp}(1)=2.981$. An estimate for $n^{\mathcal
H}_{\perp}(1)$ readily follows from the known result obtained based on the
six-loop pseudo-$\varepsilon$-expansion $n_{\mathrm{c}}=2.862$
\cite{Folk00}. Subtraction of 1 leads to the following result $n^{\mathcal
H}_{\perp}(1)=1.862$.

\section{Conclusion}\label{V}

In the present paper we have studied the conditions under which
different types of multicritical behaviour are realized for the
$O(n_{\|})\oplus O(n_{\perp})$ model. These types are related to the
three FPs ($\mathcal H$, $\mathcal D$,  $\mathcal B$), and  their
stability defines the regions in the space of the  dimensions of the OPs as well as in the
spatial dimension where the corresponding multicritical behavior
manifests itself. Using the $\varepsilon$-expansion for the two-loop
$\beta$-functions obtained in the minimal subtraction scheme we
derived the BSs separating these regions.  We obtained similar BSs
applying the resummation procedure. In the particular case of
$O(1)\oplus O(2)$ symmetry, we confirm the previous  studies finding that
the biconical  FP associated  with a tetracritical behaviour is stable
for the case $d=3$.  In higher space dimensions, the
$O(n_{\|}+n_{\perp})$ symmetrical FP associated with the bicritical
behaviour is stable.

Our analysis also made use of the results of higher order approximations
within the field-theoretical RG approach. At this stage, there were used the scaling
arguments connecting the stability of the FPs of $O(n_{\|})\oplus
O(n_{\perp})$  model with the universal quantities of the $O(n)$ and
the cubic models. Exploiting five-loop expressions for the $O(n)$
model, we derived an $\varepsilon$-expansion for the marginal dimension
$n^{\mathcal{D}}(n_{\|},\varepsilon)$ separating the regions of
stability for the FPs  $\mathcal D$ and $\mathcal B$. Applying the
resummation procedure to this result, we have analyzed the dependence
of $n^{\mathcal{D}}(1,\varepsilon)$ on $\varepsilon$. Exploiting the
five-loop expressions for the cubic model we obtained the value of
$n^{\mathcal{H}}(1,\varepsilon)$ separating the regions of stability for
the FPs  $\mathcal H$ and $\mathcal B$.  Finally, we complete our
results by three-dimensional estimates  of $n^{\mathcal{D}}(1)$ and
$n^{\mathcal{H}}(1)$ based on the pseudo-$\varepsilon$
expansions derived within a six-loop RG approximation.

These results are  also important for the critical dynamics
\cite{Folk-dynamics,Folk-dynamics_2,Folk-dynamics_3,Folk-dynamics_4,Folk-dynamics_5}. The type of a dynamical FP in such systems
depend, of course, on the static FP values. In order to extend our
results to the dynamics of antiferromagnets in an external field,
further work is necessary. One has to extend this analysis to the
statics of the corresponding model C \cite{statics_C,statics_C_2,statics_C_3}.

\section*{Acknowledgement}
This work was supported in part by the FP7 EU IRSES project N269139
``Dynamics and Cooperative Phenomena in Complex Physical and Biological
Media''.

%\end{document}
%
%% If you have problems with typesetting in ukrainian uncomment lines below.
%

\ukrainianpart

\title{Граничні вимірності для мультикритичних фазових переходів}
\author{М. Дудка\refaddr{label1},
        Р. Фольк\refaddr{label2}, Ю. Головач\refaddr{label1}, Г. Мозер\refaddr{label3}}
\addresses{
\addr{label1} Інститут фізики конденсованих систем НАН України, вул. Свєнціцького, 1, 79011 Львів, Україна
\addr{label2} Інститут теоретичної фізики Університету Йогана Кеплера, A--4040 Лінц, Австрія
\addr{label3} Інститут  фізики та біофізики Університету, A--5020 Зальцбург, Австрія}

\makeukrtitle

\begin{abstract}
\tolerance=3000%
Розглядається теоретико-польова модель, що описує мультикритичні явища і має два зв'язані параметри порядку з $n_{||}$ і $n_\perp$ компонентами
та  $O(n_{\|}){\oplus} O({n_{\perp}}\!)$ симетрію. У рамках теоретико-польової ренормгрупи вивчаються умови реалізації різних типів мультикритичної поведінки. Використовуючи двопетлеві ренормгрупові функції, розраховуються поверхні, що розділяють області стійкості для певних типів критичної поведінки, в параметричному просторі вимірностей параметрів порядку та просторової вимірності $d$.
Використовуючи розклади для ізотропної та кубічної моделей, відомі у високих порядках теорії збурень, отримуються ряди для граничної вимірності  параметра порядку, яка контролює кросовер між різними класами універсальності, до четвертого порядку за $\varepsilon=4-d$ та до п'ятого порядку
за псевдо-$\varepsilon$ параметром. Особлива увага приділяється випадку $O(1){\oplus} O(2)$  симетричної моделі, яка властива для опису анізотропних антиферомагнетиків у зовнішньому магнітному полі.
\keywords мультикритичні явища, граничні вимірності, ренормгрупа

\end{abstract}

\end{document}